\documentclass{CUPLStd}

\usepackage{amsmath}
\usepackage[T1]{fontenc}
\usepackage{lmodern}
\usepackage{color}
\usepackage{icomma}

\def\yr{\textrm{yr}}

\def\Myr{\textrm{Myr}}
\def\Gyr{\textrm{Gyr}}

\def\um{\mu\textrm{m}}

\def\cm{\textrm{cm}}

\def\pc{\textrm{pc}}
\def\kpc{\textrm{kpc}}

\def\Kelv{\textrm{K}}

\def\kms{\textrm{km\ s}^{-1}}
\def\gcm2{\textrm{g}\,\textrm{cm}^{-2}}

\def\Msun{\textrm{M}_{\odot}}

\def\AU{\textrm{AU}}

\def\nstar{n_{\star}}
\def\rhostar{\rho_{\star}}
\def\rhogas{\rho_{\rm gas}}
\def\vstar{\sigma_{\star}}
\def\vgas{\sigma_{\rm gas}}
\def\Mgas{M_{\rm gas}}
\def\Mstar{M_{\star}}
\def\SSFR{\textrm{sSFR}}
\def\SFR{\textrm{SFR}}

\def\Reff{r_{1/2}}
\def\louter{\ell_{\rm outer}}
\def\fgas{f_{\rm gas}}

\def\Htwo{H$_2$}

\def\ga{\gtrsim}
\def\la{\lesssim}

\def\endash{\text{--}}

\def\footStar{\ensuremath{\star}}
\def\nothing{\ensuremath{\varnothing}}
\def\zftn{$^{\footStar}$}

\newcommand{\BigPlus}{
  \begingroup
  \setlength{\unitlength}{0.67em}
  \linethickness{.125em}
  \begin{picture}(1,1)
  \put(0,0.5){\line(1,0){1}}
  \put(0.5,0){\line(0,1){1}}
  \end{picture}
  \endgroup
}
\newcommand{\BigMinus}{
  \begingroup
  \setlength{\unitlength}{0.67em}
  \linethickness{.125em}
  \begin{picture}(1,1)
  \put(0,0.5){\line(1,0){1}}
  \end{picture}
  \endgroup
}

\def\boldPlus{\BigPlus{}}
\def\boldMinus{\BigMinus{}}

\def\aap{{Astronomy \& Astrophysics}}
\def\aapr{{The Astronomy and Astrophysics Review}}
\def\aaps{{Astronomy \& Astrophysics Supplement}}
\def\aj{{Astronomical Journal}}
\def\apj{{Astrophysical Journal}}
\def\apjl{{Astrophysical Journal Letters}}
\def\apjs{{Astrophysical Journal Supplement}}
\def\apss{{Astrophysics \& Space Science}}
\def\araa{{Annual Review of Astronomy and Astrophysics}}
\def\icarus{{Icarus}}
\def\mnras{{Monthly Notices of the Royal Astronomical Society}}
\def\nat{{Nature}}
\def\pasp{{Publications of the Astronomical Society of the Pacific}}
\def\physrep{{Physics Reports}}
\def\prd{{Physical Review D}}
\def\qjras{{Quarterly Journal of the Royal Astronomical Society}}
\def\sovast{{Soviet Astronomy}}
\def\ssr{{Space Science Reviews}}

\newcommand{\mean}[1]{\ensuremath{\langle #1 \rangle}}
\newcommand{\reftablelabel}[1]{(\textbf{#1})}

\def\editOne{}

\jname{International Journal of Astrobiology}%

\begin{document}

\supertitle{Research Paper}

\title[Galactic Traversability]{Galactic Traversability: A New Concept for Extragalactic SETI}

\author[Lacki]{Brian C. Lacki$^{1}$}

\corres{\name{Brian C. Lacki} \email{astrobrianlacki@gmail.com}}

\address{\auadd{1}{Breakthrough Listen, Astronomy Department, University of California, Berkeley, CA, USA}}

\begin{abstract}
Interstellar travel in the Milky Way is commonly thought to be a long and dangerous enterprise, but are all galaxies so hazardous?  I introduce the concept of \emph{galactic traversability} to address this question.  Stellar populations are one factor in traversability, with higher stellar densities and velocity dispersions aiding rapid spread across a galaxy.  The interstellar medium (ISM) is another factor, as gas, dust grains, and cosmic rays (CRs) all pose hazards to starfarers.  I review the current understanding of these components in different types of galaxies, and conclude that red quiescent galaxies without star formation have favorable traversability.  Compact elliptical galaxies and globular clusters could be ``super-traversable'', because stars are packed tightly together and there are minimal ISM hazards.  Overall, if the ISM is the major hindrance to interstellar travel, galactic traversability increases with cosmic time as gas fractions and star formation decline.  Traversability is a consideration in extragalactic surveys for the search for extraterrestrial intelligence (SETI).
\end{abstract}

\received{25 May 2021}

\accepted{27 August 2021}


\Abbreviations{CR: cosmic ray, ETG: early-type galaxy, ETI: extraterrestrial intelligence, ICL: intracluster light, ICM: intracluster medium, ISM: interstellar medium, LAE: Lyman Alpha Emitter, RAIR: ram-augmented interstellar rocket, SETI: Search for Extraterrestrial Intelligence, SFG: star-forming galaxy, SFH: star-formation history, SFR: star-formation rate, $\SSFR$: specific star-formation rate, ULIRG: ultraluminous infrared galaxy}

\maketitle

\section{Introduction}

The possibility of interstellar travel looms large in the Search for Extraterrestrial Intelligence (SETI).  The classic assumption has been that extraterrestrial intelligences (ETIs) remain on their homeworlds.  Then the spatial distribution of technosignatures should closely follow the distribution of habitable worlds, and current constraints on their prevalence remain weak \citep{Morrison15}.  Sufficiently easy interstellar \editOne{migration} allows for ETIs to spread out from their homeworld to nearby stars, and then for this process to repeat at each of the secondary worlds and so on.  The first intelligence in the Galaxy could then spread to all stars within a time estimated to be $\la 100\ \Myr$ \citep{Hart75,Jones76,Newman81,Wright14-Paradox,Zackrisson15}.  The lack of \editOne{obvious} evidence for any visitation to our Solar System over its entire lifespan is commonly known as the Fermi Paradox \citep{Cirkovic18}.  Extrapolating even further, it may be possible for ETIs to modify astrophysical systems, most spectacularly by capturing their entire energy output as with Dyson spheres \citep{Dyson60,Kardashev64}.  These structures can be detected at great distances -- hundreds of parsecs for ``Type II'' shrouded stars and tens of megaparsecs for ``Type III'' shrouded galaxies (as in \citealt{Kardashev64}).  Without interstellar travel, Type III societies are impossible; with it, they might be a natural end result.  Yet observation indicates these ETIs are extremely rare \citep[e.g.,][]{Annis99,Griffith15,Villarroel16}, implying either the capability, motivation, or prospective builders themselves are missing.

\setlength\parindent{1em}
One argument against large-scale interstellar migrations is the numerous hazards on these long journeys.  Internal hazards arise from the difficulty of maintaining a stable environment for the entire trip, especially for \editOne{slow} ``world ships'' carrying large populations of organic beings from one star to another with their supporting ecologies \citep{Hodges85,Gertz20}.  Endpoint hazards prevent ETIs from permanently inhabiting a destination when they arrive, such as nearby exoplanets being uninhabitable or extant biospheres preventing migration \citep{CarrollNellenback19}.  Finally, external hazards are astrophysical phenomena that can either damage a craft or harm its population during transit \citep{Hoang17-ISM}.  Of course, susceptibility to all these hazards will depend on many unknown technologies: artificial intelligences, automated interstellar scout probes, and biological modification could all allow for easier adaptation to the journey or the destination.

\Fpagebreak

Broadly speaking, though, internal and endpoint hazards are regulated by the local stellar population and external hazards are regulated by the interstellar medium (ISM).  Dense stellar populations shorten travel times and bring more potential destinations into range \citep[as in][]{DiStefano16}.  Dense interstellar material, on the other hand, can impede a craft, and cosmic rays (CRs) can harm it or its occupants.  These factors are measured to some extent in the Solar neighborhood \citep{Frisch11,Crawford11}.  But the Solar neighborhood does not reflect the general environment of ETIs in the Universe as a whole, although it may serve as a guide for our future.  More than half the Universe's stars are in red galaxies with a very different ISM and little star formation, for example.

This paper introduces the concept of \emph{galactic traversability}, the relation between the galactic-scale environment and the ease of interstellar travel.  It is analogous to galactic habitability, which relates galactic environments and the propensity of (complex) life.  Galactic habitability has focused on the prevalence of heavy elements regulating the number of planets, radiation hazards from energetic transients like supernovae, and to a lesser extent, hazards related to flybys of stars \citep{Clarke81,Gonzalez01,Lineweaver04,JimenezTorres13}.  These factors do not necessarily promote traversability and habitability in tandem -- while high star-formation rates are correlated with frequent sterilization events and ISM hazards, low metallicity leaves fewer planets for ETIs to evolve on but also less dust to impede interstellar travel.  Galaxies of certain types may be ``super-traversable'': interstellar travel is much easier than in the Milky Way.  Traversability is a later rate-limiting step for the establishment of galaxy-wide societies.  It may be finally succeeded by \emph{galactic harnessability}, the suitability of a galaxy (and other phenomena) to being re-engineered to form a Type III entity.

\section{Components of galactic traversability}
\label{sec:Components}
The environment shapes traversability through a number of factors, listed in Table~\ref{table:Factors}.  These factors can also affect habitability in the ways summarized by the table. Factors that hinder habitability can aid traversability.

\editOne{Which aspects of a galaxy affects traversability, if any, depends on the technologies the hypothetical ETIs (or future humans) use. We have limited knowledge of the \editOne{relevant} economic and engineering trade-offs. As a rough guide, I primarily consider three qualitatively different classes of interstellar craft: (1) small (S) craft like light sails that have little mass, moving at tens of percent of $c$, like Breakthrough Starshot \citep{Parkin18}; (2) medium (M) craft, with masses perhaps in the tons to megaton range and moderate shielding, powered by nuclear fusion and achieving speeds of a few percent of $c$, like the Orion or Daedulus concepts \citep{Dyson69,Crawford90}; and (3) large (L) craft in the form of ``world ships'', limited to hundreds of kilometers per second because of the tremendous energy that must be generated and dissipated and with journey times of thousands of years. Which hazard affects which type of craft is given in brackets in the table.  A speculative class of hyperrelativistic craft (R) with Lorentz factors $\gg 1$ is also included for completeness, although attaining such speeds is difficult even with laser sails or antimatter rockets. I consider small craft to have no protection from external hazards and large craft to be basically immune to them, as the required shielding mass is prohibitively large for the former and minuscule for the latter.}

\begin{table*}[!ht]
\processtable{Factors going into galactic habitability and traversability \label{table:Factors}}
{\begin{tabular}{lcc}
\rowcolor{Theadcolor}
Quantity                     & Habitability effects                 & Traversability effects \\
\hline
Stellar density              & $\begin{array}{cp{5.5cm}} \boldMinus & \textbf{More nearby sterilization events} \\
                                                         \boldMinus & \textrm{More frequent stellar flybys that disrupt planetary systems} \end{array}$ 
                             & $\begin{array}{cp{5.5cm}} \boldPlus  & \textbf{\editOne{[All]} More destinations within range} \\
														                             \boldPlus  & \textrm{\editOne{[L]} More frequent stellar flybys bringing new destinations within range} \end{array}$ \\
\hline
Stellar velocity dispersion  & $\begin{array}{cp{5.5cm}} \boldPlus  & \textrm{Stellar flybys have weaker effects -- fewer dangerous encounters despite higher flyby rate at fixed distance} \end{array}$ 
                             & $\begin{array}{cp{5.5cm}} \boldPlus  & \textrm{\editOne{[L]} More frequent stellar flybys bringing new destinations within range} \\
                                                         \boldPlus  & \textbf{\editOne{[All] Percolation barriers dispersed quicker}} \\
                                                         \boldMinus & \textbf{\editOne{[L]} Larger minimum $\Delta v$ for destination rendezvous}\\
                                                         \boldPlus  & \textbf{\editOne{[L]} Faster travel to approaching destinations} \\
                                                         \boldPlus  & \textrm{\editOne{[M/L?]} \editOne{More powerful} gravitational slingshots} \end{array}$ \\
\hline
Specific star-formation rate & $\begin{array}{cp{5.5cm}} \boldMinus & \textbf{More sterilization events} \end{array}$
                             & $\begin{array}{cp{5.5cm}} \boldMinus & \textbf{\editOne{[S?/M/R]} Greater CR hazard} \\
                                                         \boldPlus  & \textrm{\editOne{[S]} More bright young stars for light sails} \\
                                                         \boldPlus  & \textrm{\editOne{[S]} More ISM turbulence opens rarefied corridors} \end{array}$ \\
\hline																												
Stellar metallicity          & $\begin{array}{cp{5.5cm}} \boldPlus  & \textbf{More planets for life to arise on} \end{array}$
                             & $\begin{array}{cp{5.5cm}} \boldPlus  & \textbf{\editOne{[All]} More possible planets as destination} \\
                                                         \boldPlus  & \textbf{\editOne{[All]} More minor bodies at destination to mine} \end{array}$ \\
\hline																												
\editOne{Stellar IMF}        & $\begin{array}{cp{5.5cm}} \editOne{\boldMinus} & \textrm{\editOne{More sterilization events in top-heavy environments}} \end{array}$
                             & $\begin{array}{cp{5.5cm}} \editOne{\boldPlus}  & \textrm{\editOne{[All] More potential stopover points in bottom-heavy environments}} \\
                                                         \editOne{\boldMinus} & \textrm{\editOne{[S?/M/R] Greater CR hazard in top-heavy environments}} \\
                                                         \editOne{\boldPlus}  & \textrm{\editOne{[S] More bright young stars for light sails in top-heavy environments}} \end{array}$ \\
\hline
ISM density                  & ...
                             & $\begin{array}{cp{5.5cm}} \boldMinus & \textrm{\editOne{[S/R]} Greater gas hazards}\\
                                                         \boldMinus & \textbf{\editOne{[S/R]} Higher dust density}\\
																												 \boldMinus & \textrm{\editOne{[S/R]} Larger dust grains}\\
                                                         \boldPlus  & \textrm{\editOne{[M]} More propellant for RAIR}\\
                                                         \boldPlus  & \textrm{\editOne{[S/M]} More effective magnetic sails (if ionized)} \end{array}$\\
\hline
ISM metallicity              & ...
                             & $\begin{array}{cp{5.5cm}} \boldMinus & \textbf{\editOne{[S/R]} Greater ISM dust hazard} \end{array}$\\
\hline
Hot ISM phase                & ...
                             & $\begin{array}{cp{5.5cm}} \boldPlus  & \textrm{\editOne{[S/R]} Sputtering destroys smaller dust grains}\\
                                                         \boldPlus  & \textrm{\editOne{[S]} Ionization allows magnetic sail use} \end{array}$\\
\hline
\editOne{ISM magnetic field} & \editOne{...}
                             & $\begin{array}{cp{5.5cm}} \editOne{\boldMinus} & \textrm{\editOne{[S]} Low mass craft deflected off trajectory} \end{array}$\\
\hline
\editOne{ISM radiation field} & $\begin{array}{cp{5.5cm}} \editOne{\boldMinus} & \textrm{\editOne{Fewer planets in most extreme environments?}} \end{array}$
                              & $\begin{array}{cp{5.5cm}} \editOne{\boldMinus} & \textrm{\editOne{[All] Fewer destination planets in most extreme environments?}}\\ 
															                           \editOne{\boldMinus}  & \textrm{\editOne{[R] Drag on highly relativistic craft}}\\
                                                         \editOne{\boldMinus}  & \textrm{\editOne{[R] Highly relativistic craft damaged by radiation}} \end{array}$
\end{tabular}}
{\begin{tablenotes}
A $\boldPlus$ indicates an effect that promotes habitability or traversability; $\boldMinus$ indicates an effect that hinders habitability or traversability.  The effects listed in bold most influence my judgement on whether a factor is favorable.  \editOne{The broad class of interstellar vehicles affected by each traversability effect is listed in brackets.}
\end{tablenotes}}
\end{table*}

\subsection{Stellar populations}
\label{sec:StellarPopulations}

The first major prong of traversability is the stellar population of a galaxy.  The stellar population limits the number and types of destination within a given range.  It is thus particularly important for setting the scale of internal and endpoint hazards, but it can also play some role in the journey itself.  The stellar population traits that bear on traversability include:
\begin{itemize}
\item \emph{Space density} -- The space density $\nstar$ of stars is directly proportional to the expected number of destinations within a given range at any time \citep{DiStefano16,CarrollNellenback19}.  The travel time to the nearest potential destination should, all other things being equal, likewise also decrease with travel time ($t \propto \nstar^{-1/3}$).  Patient ETIs can also wait for stars to come near their current world to shorten travel time, and the rate of these encounters increases with stellar density ($\Gamma \propto \nstar^{1/2}$; \citealt{Hansen21}).  \editOne{The space density of stars can compound other traversability hazards -- a shorter flight can mean less exposure to cosmic radiation, for example.} Overall regions with high stellar density should be more traversable.  However, high stellar density might be detrimental to habitability for analogous reasons: there are more energetic sterilizing events within a given volume and stellar encounters potentially could wreak havoc on planetary systems (\citealt{Lineweaver04,JimenezTorres13}; but see \citealt{Gowanlock11} on the limited effect of supernovae).

\item \emph{Stellar kinematics} -- All stellar systems are at least partly supported against gravity by the random velocities of stars, characterized by a velocity dispersion $\vstar$.\footnote{In addition to random velocity dispersion, the stellar populations of most galaxies have an ordered rotational component \citep{Emsellem11}.  Over distances much smaller than the system size, rotation will result in little relative motion \editOne{between nearby stars}.}  The random motions can aid or hinder stellar travel.  First, high $\vstar$ can aid travel by increasing the rate that stars come close to an inhabited system ($\Gamma \propto \vstar^{1/2}$) for patient ETIs that wait for stars to come to them.  Second, high $\vstar$ can aid the spread of ETI by mixing stellar populations.  \citet{Landis98} proposed a percolation model where only systems at the ``edge'' of inhabited space will be in range of uninhabited systems, but some of these are permanently uninterested in migration (or perhaps are by chance too far from any suitable systems).  Large regions of the Galaxy can then end up uninhabited, blocked by these ``no-go'' systems.  Stellar mixing breaks up these walls, bringing societies interested in spreading to the ``surface'' of the settlement front \editOne{\citep{Wright14-Paradox}}.  Third, $\vstar$ sets a floor on the typical $\Delta v$ between origin and destination.  High random velocities may inhibit interstellar travel, because the ETIs need to slow to a stop at the destination.  If $\vstar \sim 100\ \kms$, for example, patient ETIs using chemical propulsion will not be able to rendezvous with most destinations.  On the other hand, ETIs already using high-$\Delta v$ propulsion could save propellant by going to systems that are already approaching at high speed and simply decelerating relative to the destination the whole way.  

A speculative fourth effect is that ETIs could use stellar flybys to perform gravitational slingshots and gain $\Delta v$ for free. Significant boosts require very close approaches to large, compact masses like stellar remnants, since $\Delta v$ is limited to be around the escape velocity at closest approach.  Ancient neutron stars do in fact have a very high velocity dispersion (hundreds of $\kms$) resulting from the supernovae that birth them \citep{FaucherGiguere06}.  ETIs might use them to accelerate craft up to $\sim 1,000\ \kms$ and cross galactic distances over a few million years, if they can withstand the tides and radiation environment a few hundred thousand kilometers from the remnant.  Neutron stars have a lower space density than stars, however -- in the Solar neighborhood, the nearest is expected to be $\sim 10\ \pc$ away \citep{Ofek09} -- so the ETIs need to be able to cross the larger distances first.  The gravitational potential of a galaxy limits the $\vstar$ before stellar objects escape: many neutron stars escape their galaxies, particularly small ones.

\item \emph{Star-formation history} -- Some astrophysical phenomena are associated with young, massive stars, and their abundance is limited by the star formation history (SFH) of the stellar population.  These include O and B dwarf stars, supergiants, supernovae, bright pulsars, high-mass X-ray binaries, and gamma-ray bursts.  On the whole these phenomena may be considered dangerous to both inhabited worlds and interstellar travel because of the radiation they produce, particularly cosmic rays (see next section).  It is possible, however, that these same phenomena may draw ETIs interested in harnessing their power.  For example, they might be used to accelerate light sails to high speeds \citep{Heller17,Lingam20}.  The prevalence of these phenomena relative to old stars is encapsulated by the specific star formation rate $\SSFR \equiv \SFR / \Mstar$, where $\SFR$ is the galaxy's star-formation rate and $\Mstar$ is its stellar mass.  The stellar populations of galaxies with high $\SSFR$ have more young stars. Some energetic phenomena, like low mass X-ray binaries, are present even in older stellar populations.

\item \emph{Other properties} -- Other characteristics of the stellar population may affect traversability.  Stellar metallicity regulates the number and size of planets around stars, and perhaps the number of destinations (although terrestrial planets are not as affected as giant planets; \citealt{Buchhave14}).  It also probably limits the mass in asteroids and other minor bodies \citep{Gaspar16} that might be mined and used to build more craft.  The initial mass function \editOne{(IMF)} sets the ratio of low mass stars like the Sun to high mass stars\editOne{, and is usually similar to the local Milky Way's \citep{Bastian10}.  Locations with top-heavy IMFs, which may include the inner parsec of our Galaxy \citep{Lu13}, have an overabundance of massive stars and all their associated phenomena.  Bottom-heavy IMFs seem to exist within the inner regions of massive elliptical galaxies \citep{Smith20}, with more red dwarfs and possibly substellar brown dwarfs and sub-brown dwarfs that do not contribute much to the galaxy's luminosity.  Although habitable planets may be scarce around these objects, they may be convenient stopover points to restock and refuel, or have resources for building new structures.}
\end{itemize}

\subsection{The interstellar medium \label{sec:ISM}}
The ISM fills the space a starship must travel through, and has largely been viewed as an external hazard to interstellar travel. \editOne{Small, fast vessels like laser sails are much more vulnerable to these hazards than massive ``world ships'', although cosmic rays are sufficiently penetrating to be a hazard to smaller crewed vessels without heavy shielding.}  

\editOne{A} dynamic and complex \editOne{medium}, \editOne{the ISM has} many components that can bear on traversability:

\begin{itemize}
\item \emph{Gas} -- The great majority of the ISM's mass is in the form of gas.\footnote{Here, taken to include ionized gas (plasma).} Gas impacting a craft could harm it, with the hazards growing with speed.  Gas heats a craft; beyond a certain speed, it may cause important structures to melt and damage the surface materials \citep{Hoang17-ISM}. At relativistic speeds, ISM particles are essentially high energy cosmic rays in the craft frame, inducing radiation damage \citep{Semyonov09}.  The resulting ionization may result in a charge gain, to the detriment of small craft \citep{Hoang17-EM}.  Potentially, it could also induce drag, although gas probably does not couple efficiently with thin light sails and the drag is too weak to slow large ``world ships'' \citep{Hoang17-Drag}.  Gas could also be an opportunity for ETIs, however\editOne{, as material to be harvested for use in propulsion \citep{Bussard60}}.  The \editOne{most plausible concept in the local ISM is the} ram-augmented interstellar rocket (RAIR)\editOne{, which uses the ISM for propellant} \citep{Crawford90}.  Ionized gas might also be used as a ``wind'' to brake small craft with a magnetic or electric sail \citep{Zubrin91,Perakis16}. \editOne{Finally, a dense, highly pressurized ISM can crush the stellar wind astrospheres around stars, while the astrospheres should be larger in galaxies with rarefied ISMs.  This could affect some magnetic or electric sail designs.}

\item \emph{Dust} -- Dust grains represent $\la 1\%$ of the ISM mass \citep{Draine07-MDust}, but they could be a major hazard to interstellar travel. Each dust grain is effectively a hypervelocity micrometeoroid.  When they hit, the kinetic energy is concentrated in one spot, carving out small craters in the ship surface and causing erosion \citep{Hoang17-ISM}. At relativistic speeds, dust grains act like clusters of cosmic rays, and their impacts trigger enormous particle cascades.

The prevalence of dust varies between and within galaxies.  Dust grains require the presence of heavy elements, and dust is more prevalent in high metallicity galaxies \citep{RemyRumer14}. Generally, the density of dust is expected to track the gas density \citep{Boulanger96}.  On small scales (a few parsecs or less), though, the dust/gas ratio has been observed to fluctuate without correlation to the gas density itself \citep{Hopkins16}. Hot ionized plasma sputters away small dust grains over millions of years, reducing their threat \citep{Draine11}, although very big dust grains could survive.  

Finally, the size distribution of dust grains affects the nature of the threat.  Small grains provide a steady background of tiny impacts, whereas rarer large grains concentrate their energy into potentially catastrophic impacts.  Within the Milky Way, more mass is contained in the larger grains, up to a size of $\sim 0.2\ \um$, but few grains are $\ga 1\ \um$ wide \citep{Draine11}.  Empirical and theoretical modeling modeling suggests the size distribution of grains varies, with smaller grains more prevalent between older stellar populations and the larger sub-micron grains more prevalent in dense ISM \citep[e.g.,][]{Cardelli89,Asano13}.  A worrying prospect is that there is an undetectable ``tail'' in the size distribution of much larger dust grains that does not contribute to interstellar extinction or infrared emission, our main ways of constraining dust \citep{Hoang17-ISM}: measurements by space probes demonstrate micron-size grains exist and radar and optical searches for still larger grains have been carried out \citep{Musci12,Kruger15}.

\item \emph{Cosmic rays} -- CRs are energetic particles, mostly protons and helium nuclei, that pervade the entire ISM.  The bulk of the cosmic ray energy density is in protons with kinetic energies of order a few GeV.  There may also be regions of the ISM with significant MeV components to the cosmic ray flux, although these would be localized \citep{Indriolo09,Cummings16}.  CRs are dangerous, especially to organic beings, because of the ionizing damage they inflict \citep{Semyonov09}.  They may also damage electronic equipment.  Low energy CRs lose their energy quickly in matter by ionizing surrounding atoms, essentially searing the skin of the craft.  Higher energy CRs penetrate deeper, generating secondary particles like muons, electrons, positrons, and gamma-rays as they go \citep{Longair92}.  In the Solar neighborhood, the equivalent of several meters of water are necessary to reduce the ionizing flux from GeV CRs to tolerable levels for humans.  Alternatively, the inhabitants and mechanisms of an interstellar craft must be able to self-heal from or tolerate this radiation damage.  They may be a minor issue for thin light sails, which are mostly transparent to these particles (at the cost of being more vulnerable to gas and dust damage, as in \citealt{Hoang17-ISM}), and vast world ships where the surface shielding is a small fraction of the mass compared to the large inner volume.

\item \editOne{\emph{Magnetic fields} -- Magnetic fields thread the gases between the stars. Their traversability effects mainly apply to lightweight fast vehicles like laser sails, which become charged as they pass through the ISM.  Lorentz forces may deflect a Starshot-like sail off course by about $0.1\ \AU\ \pc^{-3}$ in the Solar neighborhood and cause it to wobble \citep{Hoang17-EM}. Very large deflections could make it difficult to deliver a probe or inscribed message into the habitable zone of fainter stars without onboard guidance. Galactic magnetic fields can include an ordered field with large-scale patterns and a fluctuating turbulent field that varies with space and time. Magnetic fields can be constrained on large scales through radio polarization and Faraday rotation mapping \citep[e.g.,][]{Beck15}, but measuring the time-variable field along a particular route may require either regular traversals or setting up a radio beacon at each terminus to provide for Faraday rotation measurements.}

\item \editOne{\emph{Interstellar radiation fields} -- The interstellar radiation field (ISRF) is dominated by near-infrared, optical, and ultraviolet light from the entire stellar population; mid- and far-infrared emission from dust in star-forming regions; and the cosmic microwave background \citep[e.g.,][]{Draine11}. Both the density and energy of photons encountered by a vessel increases with speed due to relativistic effects, and highly relativistic craft may be subject to ionization damage, radiative drag, and even nuclear reactions \citep{Hoang15,Yurtsever18}.  However, these effects will generally only be a problem at extreme Lorentz factors ($\gamma \ga 10$) well beyond those generally expected for interstellar craft. Otherwise, the effects are minimal because the ISRF is very dilute: of order $10^{-7}$ of the energy density of sunlight at Earth through most of the Milky Way.  In the most extreme starbursts, the ISRF could reach as much as $\sim 1\%$ of solar insolation on Earth, but virtually all of it is far-infrared emission which becomes ionizing only at $\gamma \ga 100$.  Likewise, a spherical craft heats to temperatures $\ga 1,000\ \Kelv$ only if $\gamma \ga 100$ even in extreme starbursts.\footnote{\editOne{For craft with $v/c \approx 1$, average received photon energy $\propto \gamma$ and average craft-frame ISRF energy density $\propto \gamma^2$.}} A more indirect effect might be the inability of extreme starburst environments to form planets to go to \citep{Thompson13}, although pre-existing planets would be unaffected. It is unclear how the changes to planet formation affect the prevalence of minor bodies.}

\end{itemize}

\section{A broad view of galactic traversability}

The global properties of galaxies, their stellar populations, and their ISM interplay with each other and with the surrounding intergalactic medium.  A picture of how galaxies evolve has emerged over the past couple of decades, and with it, we can gain an overview of galactic traversability through the Universe.

\subsection{Red and blue: Galaxy evolution and trends in traversability}
\label{sec:RedVsBlue}
Most galaxies fall into two broad categories, the blue star-forming galaxies (SFGs) and the red quiescent galaxies \citep{Strateva01}.  At $z \sim 0$, the quiescent galaxies include most elliptical and lenticular (S0) galaxies and are biased towards dense galaxy clusters, while SFGs include most late-type spiral and irregular galaxies and are typically found in the field \citep{Skibba09}.  These groups differ fundamentally in their $\SSFR$. The sSFR of blue galaxies with a given stellar mass $\Mstar$ are similar at any given cosmic epoch \citep{Brinchmann04}, evoking a SFR--$\Mstar$ relation sometimes called the ``main sequence'' \citep{Noeske07,Renzini15}.\footnote{Not related to the (epoch-independent) main sequence of stars.}  This characteristic $\SSFR$ has been decreasing for the past $\sim 10\ \Gyr$ \citep{Noeske07,Speagle14}.  Quiescent galaxies, by contrast, have a very small $\SSFR$ and have formed few stars over the past gigayears.  Thus, SFGs have a mixture of young and old stars, while quiescents predominantly have old, lower mass stars.  Between the red and blue galaxies lie a transitional group of ``green valley'' galaxies, which the Milky Way is entering \citep{Licquia15}, while starburst galaxies have abnormally high $\SSFR$ for their epoch \citep{Elbaz11}. A small fraction of galaxies at any given time have active galactic nuclei, with the central black holes emitting prodigious electromagnetic and particle radiation.

The dichotomy in $\SSFR$, and its cosmic history, arises from the evolution of the gas supply in galaxies, a major factor in traversability.  Star formation requires the presence of sufficient amounts of cold (generally molecular) gas that can collapse, with a larger star-formation density in regions of dense gas \citep{Kennicutt98}.  Galaxies are fed gas from the surrounding intergalactic medium, and convert some of that gas into stars at an equilibrium rate set by feedback from the young stars, while some is ejected in a wind \citep{Bouche10}.  SFGs notionally climb up the main sequence, from dwarf to giant, but the main sequence as a whole has falling $\SSFR$ with time, and blue galaxies had much higher gas fractions ($\fgas \equiv \Mgas / (\Mgas + \Mstar)$) and star-formation rates in the early Universe \citep{Santini14,Tacconi18}.  A large main sequence SFG with $\Mstar \sim 10^{11}\ \Msun$ in the present day can be expected to have a gas mass $\Mgas \sim 10^{10}\ \Msun$ and $\SFR \la 10\ \Msun\,\yr^{-1}$, but a similarly sized galaxy ten billion years ago would have had $\Mgas \sim 10^{11}\ \Msun$ and $\SFR \ga 100\ \Msun\,\yr^{-1}$ \citep{Daddi10,Tacconi13}.  As both the star-formation rate and the gas fraction fall with time, the interstellar media of SFGs are becoming more traversable.  The generations of massive stars born and dying during this growth results in metals accumulating in the ISM, and high-$\Mstar$ galaxies have higher metallicity even if they formed all their stars rapidly in the early Universe \citep{Mannucci10}. A higher metallicity leads to more planets, but also more dust in ISM, favoring habitability at the expense of SFG traversability.

Growth in the main sequence mode does not continue forever, but is halted when the SFG runs out of gas.  This can be a sudden process, triggered by a galaxy merger funneling gas into the core to be consumed in a brief spectacular starburst, but more often is the result of a slower exhaustion or ``strangulation'' of the external gas supply that results in the galaxy crossing the green valley \citep{Schawinski14,Peng15,Bremer18}.  The building of a bulge may also be a factor in this quenching process: many green valley galaxies are early-type spirals and many quiescents appear to have stellar disks within them \citep{Bremer18}.  Within a couple billion years, the quenched former SFGs are red and quiescent.  They are broadly expected to have little gas -- and thus be favorable in terms of traversability.  However, they can retain a residue of gas from their past lives as SFGs, accrete it from passing smaller galaxies with their tides, and replenish it as old stars eject their envelopes, so their ISMs are not entirely empty \citep{Davis11,Young14}.

\subsection{Star-forming galaxies: A cold, shifting labyrinth}
\label{sec:SFGs}
The ``space climate'' of SFGs is shaped by the chaotic and violent interplay between gas and star formation.  The space weather within them is thus very chaotic, inhomogeneous and constantly changing.  On a large scale, the ISM is complicated by its ``multiphase'' nature, where each phase has a different temperature and pressure.  They all have roughly the same pressure, so hot gas is less dense \citep{Draine11}.  Classically, there are three phases to the ISM, although this is a great simplification \citep{Cox05}.  The cool/cold phase ($\sim 10 \endash 100\ \Kelv$) has dense neutral (HI) and molecular (\Htwo) gas, the warm phase ($\sim 10,000\ \Kelv$) of neutral (HI) and ionized (HII) gas, and a third hot phase ($\sim 10^6\ \Kelv$) is carved out by supernovae and young stellar clusters.  In the Milky Way, the hot phase fills about half the volume, a patchwork of interstellar space covering much of the Galactic disk where gas and dust pose little obstacle to starfarers.  Most of the remaining volume is filled with warm gas, and only small clouds and filaments with dense cold gas.\footnote{The Solar neighborhood contains examples of both the hot and warm ISM: the Local Bubble and local interstellar clouds, respectively \citep{Frisch11}.} The different phases of gas also have different height distributions, with cold gas concentrated close to the Galactic plane, while warm ionized gas rises several hundred parsecs above the plane, punctured by chimneys of hot gas over star clusters \citep{Norman89,Cox05}.  The standard three phase structure is expected to apply to most contemporary spiral galaxies, but it is not universal.  Intensely star-forming galaxies, which includes both starbursts in galactic cores and most large SFGs in the early Universe, are dominated by vast amounts of cold gas \citep{Daddi10,Tacconi18}.  A high-volume phase of hot ISM may also exist in weaker starbursts, excavated by the high rate density of supernovae, but this is not agreed upon \citep{Strickland07,Krumholz20}.  

These phases are not simple clouds or smooth media.  Turbulence is found in at least the cold and warm phases of SFGs, with $\vgas \sim 10\ \kms$ in contemporary main sequence SFGs and up to $\sim 100\ \kms$ in starbursts and high-redshift SFGs \citep[e.g.,][]{Downes98,Dib06,Law09}.  Turbulence creates random density fluctuations at a variety of length scales, the largest at the outer scale $\louter \sim 0.1 \endash 1\ \kpc$ \citep[e.g.,][]{Draine11}. The amplitude of these fluctuations grows according to the Mach number of the turbulence (ratio of $\vgas$ to gas sound speed), surpassing the mean density when the turbulence becomes supersonic.  In ISM with supersonic turbulence, most of the matter is concentrated into dense transient clumps while most of the volume is rarefied, with $\sim 10 \%$ at any given time having $\rhogas \ge \mean{\rhogas}$ \citep{Padoan97,Ostriker01,Federrath13}.  The fluctuations evolve rapidly, growing and dissipating within one flow crossing time $\louter / \vgas$ \citep{Stone98,MacLow99}, resulting in constantly shifting space weather in ISM with supersonic turbulence. The Mach number of the turbulence is much higher in cold gas due to the lower sound speed, and thus the cold, denser gas has large density contrasts, with most of the mass concentrated into transient clumps, with the densest forming molecular clouds.  Thus, even in starbursts with a mean ISM density hundreds of times higher than the Milky Way's, at any given location the gas and dust density may be relatively low.  Yet although these relatively traversable conditions are the median, they are interrupted by intervals where traversability is low.  Thus a map of the traversability of a SFG would constantly shift over millions of years.

If gas and dust truly are major impediments to interstellar travel, SFGs are essentially mazes where travel can occur freely along some paths (particularly in the hot ISM) but are walled off by dense matter along other paths.  The free paths are not necessarily connected, which can lead to a mosaic of disconnected regions in a galaxy, between which travel is impractical.  This labyrinthine patchwork does suggest that percolation barriers to interstellar migration could apply on short enough timescales.  Ultimately, however, the dynamic nature of the ISM thwarts these blockages on long enough timescales: the walls of the labyrinth are constantly moving, opening up new paths and connecting some regions even as they close off and split others.

The same cannot be said if the main hazard is cosmic rays.  Plasma waves in the ISM trap CRs and prevent them from simply flying out of a galaxy.  While GeV CRs are thought to be mainly accelerated in supernova remnants and bubbles surrounding young star clusters, they diffuse many hundreds of parsecs from these sites, when they are not destroyed by interactions with the ISM \citep{Murphy08}.  Nor are they confined to the galactic plane: isotopic evidence in the Milky Way implies they can wander up and down by at least a kiloparsec off the plane until they start to escape \citep{Lukasiak94,Strong98}.  As such, the bulk CR density is not expected to vary too much within the stellar disks of most SFGs, although in the Milky Way it slightly decreases with distance from the center \citep{Acero16}. There is no escape from the CR hazard unless one moves far into the galactic halo, potentially hindering travel between most of the stars in a SFG.  The mean background level of CRs between SFGs varies dramatically, however.  Gamma-ray and radio observations indicate that the CR energy density increases rapidly with star formation rate: it is significantly lower in dwarf irregular galaxies \citep{Abdo10-SMC}, but hundreds of times higher in starburst regions and probably large SFGs at high redshift \citep[e.g.,][]{Acciari09,Papadopoulos10}. 

\editOne{Magnetic fields are generally observed to have a nearly equal energy density as cosmic rays in local star-forming spiral galaxies, although the fields likely become dominant in the most intense starburst regions \citep[e.g.,][]{Thompson06,YoastHull16}. Their ubiquitous presence is attested by the radio synchrotron emission observed from SFGs throughout the Universe, and they may be maintained by a dynamo driven by ISM turbulence even in the early Universe \citep{Arshakian09}.}

Of course, these conclusions depend on how much of a hindrance each part of the ISM is.  It is possible that ETIs use RAIR propulsion and favor cold, dense ISM to acquire propellant -- then SFGs may instead have fleeting ``islands'' of ETI travel separated by empty gulfs instead of a maze.  Starbursts and high-redshift SFGs may then actually be favored for interstellar travel because of their dense gas reservoirs, if ETIs can cope with the extreme levels of CRs in them.  Or the reality may be more nuanced, with a mosaic of different interstellar travel methods, each adapted to presiding local conditions.

\subsection{Quiescent galaxies: Placid until they're not}
\label{sec:Quiescent}
The space climate of quiescent galaxies is very different.  While SFGs are filled with stormy and chaotic ``weather'', the ``atmosphere'' of these galaxies is generally thinner and more placid.  In some galaxies this seeming quiet can be disturbed by the wildness of star formation -- or long galaxy-wide disasters when the central black hole turns on.

Like SFGs, quiescent galaxies come in a vast ranges of mass and size, from ultrafaint dwarf spheroidal galaxies to supergiant ellipticals in the hearts of galaxy clusters.  Generally, the larger the galaxy, the larger the stellar velocity dispersion: in dwarf spheroidals, $\vstar$ can be as small as $1\ \kms$ \citep{Simon19}; in the hearts of the largest ellipticals, over $300\ \kms$ \citep{Cappellari13-Rels,Norris14}.  Moderate to large ETGs, which contain more stellar mass, have been discovered to fall into two broad categories, the fast rotators and the slow rotators \citep{Emsellem07,Cappellari16}.  Fast rotators, which include the lenticulars and the majority of ellipticals, have stellar disks like their progenitors, the spirals, although their bulge components are usually dominant \citep{Krajnovic13}.  Slow rotators are a diverse class, with some being disks or containing two counterrotating populations.  The largest slow rotators are giant elliptical galaxies with rounder ``boxy'' shapes \citep{Kormendy96}, and thus I will refer to them as boxy ETGs.  The dichotomy extends to many galaxy properties \citep{Kormendy09} and appears to have a profound effect on the ISM.

The classical expectation of quiescent galaxy ISMs is that they do not have any, but observations have shown that this is not universally true.  A significant fraction of ETGs (particularly fast rotators outside of galaxy clusters) have cool atomic and molecular gas.  Estimates of the ETG fraction with HI range from $10\%$ to $70\%$ depending on environment and the sensitivity to gas \citep{Morganti06,Serra12}, and $\sim 20\%$ have molecular gas \citep{Young11}.  In some cases, the amount of cool gas can reach levels comparable to the Milky Way, but in most cases the gas mass is much lower ($\sim 10^6 \endash 10^9\ \Msun$).  Dust is also commonly present in ETGs, if usually at much lower abundances than in normal star-forming spirals \citep{Goudfrooij95,Ferrari99}.  \citet{Smith12} detect dust in 62\% of lenticular galaxies and 24\% of ellipticals.  In those galaxies with detections, the mass ratio of dust to stars is $\sim 10^{-4}$, compared to $\sim 10^{-3}$ in the Milky Way \citep{Smith12}.  The dust is often concentrated in a small disk around the galactic nucleus or in filaments, so ETIs may simply be able to dodge what dust is present \citep{Tran01}.  This is not always the case, however \citep{Goudfrooij95,Temi07}.

Cool gas is fuel for star formation, suggesting a CR threat.  Star formation is indeed observed, usually in fast rotator galaxies, as is the radio synchrotron associated with CRs \citep{Shapiro10,Young14,Kokusho17,Nyland17}.  According to \citet{Shapiro10}, the star-formation appears to have two ``modes'', either spread throughout the galaxy or concentrated near the core.  The former only occurs after the ETG receives a windfall of cool gas from an encounter.  \citet{Shapiro10} speculates that nuclear star formation occurs in occasional spurts within all fast rotators, supplied by gas from the old stellar population.  Either way, however, star formation is a brief episode in a largely quiescent history, limiting the CR threat temporally.  Nuclear star formation also likely results in a CR threat only in the inner regions of the host galaxy.  Because there usually is less cool gas than in proper SFGs, the star-formation rate should also be lower.  There is also evidence that the star-formation efficiency -- the rate at which gas is converted into stars -- is lower in elliptical galaxies, which would suppress the SFR in these galaxies even further (\citealt{Martig13,Nyland17}; \citealt{Kokusho17} disputes this, however).  Finally, the radio-faintness of ETGs with star formation could suggest that CRs escape easily in these galaxies \citep{Nyland17}.  All of these factors would limit the CR energy density in intensity, space, and time.

Quiescent galaxies also have hot gas supplied by dying old stars.  This gas gets stirred up by the stellar population, thermalizing it to $10^6 \endash 10^7\ \Kelv$ where the sound speed is roughly the stellar velocity dispersion and further heated by the Type Ia supernovae of old white dwarfs \citep{Ciotti91,Mathews03}.  The fate of this gas depends on the depth of the galaxy's gravitational potential.  In the majority of ETGs, the hot gas is heated until it escapes, as borne out by a lack of X-ray emission \citep{Mulchaey10,Sarzi13}.  The most massive ETGs (including the boxy ETGs) hold onto their hot gas, which builds over time. Because the hot gas is glowing so brightly X-rays, it is losing energy, cooling down, and sinking towards the galactic center. Some of this gas may form stars, but the real trouble is that it provides a fuel source for the central black hole, which in turns powers relativistic jets seen in radio.  Indeed, \emph{all} massive ETGs appear to be radio galaxies on some level \citep{Brown11,Sabater19}, though most of the time the activity appears to be confined to a small core \citep{Slee94,Nyland16}.  From time to time, high accretion onto the central black hole fuels extraordinary episodes of activity in which jets blast their way through the galaxy and inflate lobes of CRs.  The CR lobes can fill much of the galaxy's volume with dangerous levels of radiation, which can be ten to hundreds of times the Galactic background \citep{McNamara07,McNamara12,deGasperin12,Pfrommer13}.  \editOne{Magnetic field strengths of tens of microgauss are estimated in these regions \citep{deGasperin12}.} Thus these larger ETGs may have long time periods where interstellar travel is relatively safe from CRs punctuated by catastrophes lasting tens of millions of years wherein radio jets and lobes make interstellar space extremely hazardous.

Another potential source of CRs in quiescent galaxies are the Type Ia supernovae in old stellar populations \citep[as in][]{Lien12}, but it is not clear how many CRs these accelerate or how the radiation propagates. \editOne{Likewise, magnetic fields in the quiescent ISM are very poorly constrained, although they probably are weaker than in the Milky Way \citep{Seta21}.}

\subsection{Role of size evolution of galaxies}
\label{sec:SizeEvol}

Galaxies grew from the inside-out, with the denser cores forming at high redshift \citep[e.g.,][]{Nelson16}. This secular growth is supplemented by galaxy mergers. Today's galaxies were more compact in the distant past, which partly offsets the lower stellar mass when considering $\rhostar$ and the typical duration of interstellar journeys. This size evolution is not solely due to \editOne{galaxies with less stellar mass being smaller}: in fact, galaxies of any fixed stellar mass were more compact in the early Universe.  At $z \sim 2$, SFGs were about half as wide as present galaxies of the same mass, implying four times the stellar surface density \citep{Trujillo06-SizeEvol,vanDerWel14}.  Note that at fixed $\Mstar$, we select different galaxies at different cosmic epochs. When following an \emph{individual} SFG, $\rhostar$ increases over time in all regions of the galaxy \citep{vanDokkum13}. Still, the typical inhabited planet of the early Universe could very well have been located in a relatively compact and massive galaxy, simply because the sparser low-$\Mstar$ galaxies had not formed many stars yet.

Despite the smaller radii, the median population-weighted $\rhostar$ probably was not much greater than at present. SFGs in the early Universe had higher levels of turbulence, which puffs out their gaseous disk and the stars formed from them \citep[e.g.,][]{Genzel08}. \citet{Wisnioski15} estimates the typical $\vgas$ of SFGs was $\sim 50\ \kms$ at $z \sim 2.3$, whereas it is now $\sim 10\ \kms$.\footnote{The stellar velocity dispersions are significantly higher, $\sim 50\ \kms$ in the Milky Way. Older stars in the Galaxy have higher velocity dispersions. This partly could be due to a ``heating'' mechanism that increases $\vstar$ over time, but the higher initial turbulence in the disk when these stars formed is another probable cause \citep{Lehnert14}.} As the disk thickness-to-radius ratio varies proportionally to $\vgas$, the more compact radial scales at high redshift were offset by greater vertical scales, with a relatively weak net effect on traversability.

Although quiescent galaxies were comparatively rare in the early Universe, they were even more compact, about one-quarter the radius of present-day quiescent galaxies with the same $\Mstar$ \citep{Trujillo06-RedNugget,vanDerWel14}. These ``red nuggets'' had $\Mstar$ of $\sim 10^{10} \endash 10^{11}\ \Msun$ concentrated into a disk of scale radius $\sim 1\ \kpc$, with $\rhostar$ hundreds of times that found in the Milky Way \citep{vanDokkum08}. Most of these galaxies have seemingly disappeared. Nonetheless, environments of similar $\rhostar$ -- and presumably traversability -- are found in the inner kiloparsecs of today's massive early-type galaxies, suggesting that some of them are red nuggets that accreted stars during galaxy mergers \citep{Bezanson09}.

\subsection{Role of intergalactic environment}
\label{sec:Environment}
Galaxies are not isolated, but are grouped hierarchically.  In particular, dwarf galaxies are frequently satellites of large galaxies, and even large galaxies are subhalos when they are located in galaxy clusters.  The parent system usually has an associated gaseous medium itself, resulting in drag on the smaller system's ISM as it moves with high speed through the surrounding gas.  Tidal forces can also dislodge ISM and stars from the smaller galaxy.  These effects deprive the galaxies of their ISM, shutting down star formation, and rendering them super-traversable.

\section{Which galaxies are traversable?}
There is a menagerie of different kinds of galaxies, and in this section I discuss the traversability trends that can be expected between them.  Table~\ref{table:GalaxyTypes} summarizes the presence of various components of galactic traversability, according to galaxy types.  The given quantities should be understood as just examples: even within each type of system listed (particularly dwarf galaxies), basic quantities like stellar mass can vary by orders of magnitude.  In addition, many of the quantities can vary by orders of magnitude \emph{within} galaxies.  The listed densities should be regarded as a loose mass-weighted average or median values, roughly the density in the neighborhood of a typical star.   

\begin{table*}[!ht]
\tabcolsep4pt
\processtable{Traversability of examples of stellar system\label{table:GalaxyTypes}}
{\begin{tabular}{lcccccccccc}
\rowcolor{Theadcolor}
Type & \multicolumn{5}{c}{Stellar population} & \multicolumn{4}{c}{ISM} & Ref \\
\rowcolor{Theadcolor}
     & $\Mstar$ & $\Reff$ & $\rhostar$        & $\vstar$ & SFH & Phases & $\rhogas$       & Dust/gas & CRs & \\
\rowcolor{Theadcolor}
     & $\Msun$  & $\kpc$  & $\Msun\,\pc^{-3}$ & $\kms$   &     &        & $m_H\,\cm^{-3}$ &          &     & \\
\hline
Solar neighborhood     						  & $10^{10.8}$         & $3$      & $0.05$      & $50$   & Mix       & All        & $10^0$          & M        & M          & 1\\
                                    &                     &          &             &        &           & Hot        & $\la 10^{-2}$   & L/M      & M          &\\
                                    &                     &          &             &        &           & Warm       & $\sim 10^{0.5}$ & M        & M          &\\
                                    &                     &          &             &        &           & Cold       & $\ga 10^1$      & M        & M          &\\
\hline
Dwarf irregular        						  & $10^8$              & $0.5$    & $10^{-1.5}$ & $25$   & Young/Mix & All        & $10^0$          & L        & L          & 2\\
Spiral                              & $10^{10}$           & $2.5$    & $10^{-1}$   & $50$   & Mix       & All        & $10^0$          & M        & M          & 3\\
Nuclear starburst                   & $\ga 10^8$          & $0.2$    & $\ga 10^0$  & $50$   & Young     & Cold       & $\ga 10^{+1.5}$ & M        & VH         & 4\\
Small high-$z$ (LAE) \zftn          & $10^9$              & $1$      & $10^{-1.5}$ & $50$   & Young/Mix & All?       & $10^0$          & L        & M/H?       & 5\\
Large high-$z$ (ULIRG)\zftn         & $10^{11}$           & $5$      & $10^{-1.5}$ & $100$  & Mix       & Cold       & $10^0$          & M        & H          & 6\\
\hline
Dwarf spheroidal (field)            & $10^7$              & $0.5$    & $10^{-2.5}$ & $20$   & Old       & Cold       & $10^{-3}$       & L        & \nothing   & 2, 7\\
\hphantom{Dwarf spheroidal} (halo)  &                     &          &             &        &           & \nothing   & \nothing        & \nothing & \nothing ? & \\
Compact elliptical                  & $10^{8.5}$          & $0.15$   & $10^{+0.5}$ & $50$   & Old       & \nothing ? & \nothing        & \nothing & \nothing   & 7\\
Fast rotator (gas-rich)             & $10^{10.5}$         & $3$      & $10^{-1.5}$ & $125$  & Old       & Cold       & $10^{-1.5}$     & M        & L          & 8, 9\\
\hphantom{Fast rotator} (gas-poor)  &                     &          &             &        &           & (Hot)      & \nothing        & \nothing & \nothing   & \\
Boxy ETGs                           & $10^{11.5}$         & $10$     & $10^{-2}$   & $225$  & Old       & Hot        & $10^{-2.5}$     & L        & L?/H       & 9, 10\\
Red nugget\zftn                     & $10^{11}$           & $1$      & $10^{+0.5}$ & $300$  & Old       & Hot        & $10^{-1}$       & L        & L?/H?      & 11\\
\hline
Globular cluster                    & $10^5$              & $0.003$  & $10^{+2}$   & $5$    & Old       & Warm       & $10^{-1}$       & L?       & \nothing ? & 7\\
Galaxy cluster (ICL/ICM)            & $10^{11.5}$         & $200$    & $10^{-6}$   & $1000$ & Old       & Very hot   & $10^{-3}$       & \nothing & L          & 12
\end{tabular}}
{\begin{tablenotes}
All values are only order of magnitude estimates.  Qualitative ratings are given relative to mean Solar neighborhood values, which is given a medium (M) rating.  Other abbreviations: \nothing = none known; L = low; H = high; VH = very high.  When densities were not found in the literature, they were calculated from stellar or gas mass and radius.  For exponential disks (in SFGs, dwarf spheroidals, and red nuggets) with radial scale length $R$ and scale height $h$, the mass-weighted mean density is $M / (32 \pi R^2 h^2)$.  I used $h = R/10$ for the spiral galaxy, and $R/4$ for the others. For spheroids (compact ellipticals, fast rotators, boxy ETGs, globular clusters, galaxy clusters), I estimated based on the \citet{Dehnen93} profile, in which the density at the half-mass radius $R$ is $\sim 0.3 M / (4 \pi R^3)$.\\
\footStar: Found mainly in early Universe.\\
\textbf{References} -- \reftablelabel{1} \citet{Cox05,Licquia15,Anguiano18}; \reftablelabel{2} \citet{McConnachie12}; \reftablelabel{3} \citet{Leroy08}; \reftablelabel{4} \citet{Kennicutt98}; \reftablelabel{5} \citet{Gawiser07,Bond09}; \reftablelabel{6} \citet{Tacconi13}; \reftablelabel{7} \citet{Norris14}; \reftablelabel{8} \citet{Young11,Serra12}; \reftablelabel{9} \citet{Cappellari13-Rels}; \reftablelabel{10}: \citet{Fukazawa06}; \reftablelabel{11}: \citet{Trujillo14}; \reftablelabel{12}: \citet{Vikhlinin06,Cavaliere13}
\end{tablenotes}}
\end{table*}

\emph{Dwarf irregular galaxies} -- the low mass end of the SFG ``main sequence'' at $z = 0$, with the Small Magellanic Cloud being among the largest.  They are smaller than spiral galaxies with higher gas fractions $\Mgas/\Mstar$, so neither the stellar population nor the ISM are necessarily much less dense than the Solar neighborhood.  Dwarf irregulars have either constant SFRs or have recently restarted star formation, after an initial episode in the very distant past \citep{Weisz14}.  These low mass galaxies have shallow gravitational potentials and lower $\vstar$, with an escape velocity low enough that virtually all neutron stars escape the galaxy.  As metal-poor galaxies, they have low dust-to-gas ratios: about one-tenth the Milky Way value for Magellanic-like irregulars to one-thousandth for the smallest galaxies, albeit with significant individual variations \citep{Lisenfeld98,RemyRumer14}.  CRs are present, but at lower levels than in the Milky Way, and they probably escape more easily \citep{Murphy08,Lacki10-FRC1,Abdo10-SMC}.  \editOne{These galaxies contain relatively weak microgauss magnetic fields \citep{Roychowdhury12}.} Because of the lower abundance of dust and CRs, these galaxies may be somewhat favorable for traversability overall, although the stellar population may be somewhat unfavorable.

\emph{Spiral galaxies} -- the medium to large ``main sequence'' SFGs of $z = 0$.  Technically, the Milky Way does not lie on the main sequence, having a somewhat low $\SSFR$ and redder color \citep{Licquia15}, but the Milky Way's properties are of the same order-of-magnitude.  Moderately higher $\fgas$ are common in smaller (generally late-type) spirals \citep{Santini14,Tacconi18}.  Dust-to-gas ratios are near the Galactic value \citep{Draine07-MDust}, although generally increasing with stellar mass \citep{RemyRumer14}.  Gamma-ray observations suggest that the CR background is less intense in M33, a small late-type spiral, than in the Milky Way \citep[e.g.,][]{Ackermann17}. \editOne{Magnetic fields in spiral disks include both a large-scale organized component and a turbulent field, both having strengths of a few microgauss, up to about twenty microgauss in parts of some galactic disks \citep{Beck15}.}

\emph{Nuclear starbursts} -- violent star-formation events in the cores of SFGs, fueled by a very high concentration of gas.  With extremely high gas densities, high ISM metallicity, and extraordinary levels of CRs, they are likely extremely unfavorable for traversability. The extreme CR levels would pose severe problems for all space travel, even close to a home planet. \editOne{ISM magnetic fields of several hundred microgauss are common, up to several milligauss strength in some cases \citep{McBride14,YoastHull16}. This can lead to deflections of Breakthrough Starshot-like laser sails of $10 \endash 100\ \AU\,\pc^{-3}$, a problem compounded by the evolving turbulent fluctuations in the field. The ISRFs are hundreds of times more intense than in the Solar neighborhood or more, with energy densities comparable to the solar insolation at Jupiter in the most extreme cases \citep[e.g.,][]{Lacki10-FRC1,Thompson13}.} The gas and dust threat may be reduced if channels of hot gas exist, but even then the mean density is expected to be $\sim 1\ m_H\,\cm^{-3}$ \citep{Strickland07}. If the ISM can be shielded against, however, these regions can have $\rhostar$ an order of magnitude or more than the Solar Neighborhood, which would favor traversability after all. Similar $\rhostar$ are found in the cores of modern quiescent galaxies, however, which are both more numerous and less hostile to life.

\emph{High-$z$ SFGs} -- SFGs in the early Universe, when gas fractions and star formation rates were much higher. Typical $\rhostar$ were probably of the same order as found in present SFGs, though with higher $\vstar$. The small high-$z$ SFG represents the ancestor of a Milky Way-like galaxy at $z \sim 2 \endash 3$, and includes typical Lyman $\alpha$ emitters (LAEs) at this cosmic epoch \citep{Guaita10}.  The mean SFR of LAEs was a few times that of the present Galaxy, but it was concentrated in the inner kiloparsec \citep{Gawiser07,Bond09}.  The early Milky Way likewise formed stars quickly \citep{Snaith15,Gallart19}.  CRs were therefore a bigger threat, although the low metallicity reduced the dust hazard.  \editOne{These galaxies hosted stronger (presumably turbulent) magnetic fields associated with their intense star formation \citep{Lacki10-FRC2}.} The properties of the large high-$z$ SFG in Table~\ref{table:GalaxyTypes} represents the ultraluminous infrared galaxies (ULIRGs) that dominated the Universe's star formation ten billion years ago.  They had ten times the gas and dust mass as the present Milky Way, although generally spread out in thicker disks, and hundreds of times the SFR and CR radiation levels \citep{Tacconi13,Tacconi18}.  These galaxies were extremely clumpy \citep{Genzel11}, so traversability probably varied considerably across a galaxy.  Overall, high-redshift SFGs likely were much less traversable, especially the larger ones.

\emph{Dwarf spheroidals} -- one of the two main types of small ETGs and by far the most numerous.  The smallest are the ultrafaint galaxies, which can have just a few thousand stars \citep{Simon19}.  Dwarf spheroidals are small, low mass, low density, and low metallicity \citep[see][]{McConnachie12}.  In the largest members of the class, like NGC 205, $\rhostar$ may typically reach $\sim 0.1\ \Msun\,\pc^{-3}$.  The numerous small dwarf spheroidals have $\rhostar$ more than order of magnitude sparser than the Solar neighborhood, adversely affecting traversability. Their gravitational potential is low enough that practically all neutron stars should have escaped long ago.  In addition, even a single supernova can unbind the ISM and blow it away, as can ram pressure stripping in a large galaxy's halo, although the ISM can be replenished by old stars.  Dwarf spheroidals near the Milky Way and M31 ($\la 250\ \kpc$) lack any detectable gas, while those far from galaxies have low density reservoirs of HI gas \citep{Blitz00,Grcevich09}.  Without star formation, they also lack CRs \editOne{and magnetic fields}, as confirmed by deep radio observations \citep{Regis15}.  Dwarf spheroidals are dwarf irregulars without the ISM hazards. If the stellar population dominates traversability, then all but the largest dwarf spheroidals are sub-traversable, but if the ISM is the main barrier to interstellar travel, they may be even more traversable than the Milky Way. But a big question is whether anyone lives in them to test this supposed advantage, or whether there is even anywhere to go, given the probable dearth of planets (and possibly minor bodies).  The smallest have \emph{mean} metallicities of $1/300$ Solar \citep{Simon19}, below the predicted $\sim 1/100$ Solar physical threshold for terrestrial planet formation \citep{Zackrisson16}.

\emph{Compact ellipticals} -- the other main type of small ETGs and relatively rare, these probably are the cores of larger ETGs that have been stripped of the rest of their stars by tides \citep{Chilingarian15}.  Their smaller cousins ($10^6 \endash 10^8\ \Msun$; $10\ \pc$) are the ultracompact dwarf galaxies \citep{Norris14}.  The stars are close together, and large ones like M32 have moderately high stellar velocity dispersions.  As the bare cores of ETGs, they could have a low density ISM supplied by their old stars.  None has been detected in M32 \citep{Sage98,Welch01,Boroson11}.  Dust should therefore be of little impact despite the high metallicity of stars.  The combination of high stellar density and nearly empty ISM is extremely favorable for interstellar travel: these are super-traversable galaxies.

\emph{Fast rotators} -- the most common type of medium to large ETGs \citep{Emsellem11}, containing a significant fraction of the Universe's stellar population.  Stellar density is strongly peaked towards the nucleus and thus ranges from very high to very low \citep[e.g.,][]{Dehnen93,Kormendy09}.  The inner kiloparsec of large fast rotators have $\rhostar$ tens of times greater than in the Solar Neighborhood, comparable to a red nugget \citep{Bezanson09}. Most stars are probably located further out, where $\rhostar$ is comparable to the Solar Neighborhood, however. All but the largest are unable to retain the hot gas from their old stars and are X-ray faint \citep{Mulchaey10,Sarzi13}.  The presence of a cool ISM depends on galaxy type, with a higher prevalence in lenticulars than ellipticals, and environment, with 40\% of fast rotators in the field but only 10\% in the Virgo Cluster having HI according to \citet{Serra12}.  Environment does not seem to affect how often they have molecular gas \citep{Serra12}.  Even in most fast rotators with a cool ISM, traversability is likely favorable because gas and star formation are less abundant than in the Milky Way.  Those without a cool ISM should be super-traversable. 

\emph{Boxy ETGs} -- the largest ETGs, often found in the centers of galaxy clusters, with stellar masses an order of magnitude bigger than the Milky Way.  Because the stars fill a round volume with scale length of several kiloparsecs instead of being squashed into a disk, the mass-averaged stellar density is relatively low, but stellar velocity dispersions are high.  Like fast rotators, stellar density varies widely from core to outskirts -- although the density does not rise as sharply in the inner regions \citep{Kormendy09}, it is nonetheless very high in the inner kiloparsec \citep{Bezanson09}.  The hot ISM reaches a peak density of $0.1\ \cm^{-3}$ in the core \citep{Mathews03} -- about ten times that of the Local Bubble.  The ISM density declines away from the center, though, and is $\la 0.01\ \cm^{-3}$ a few kiloparsecs out \citep{Fukazawa06}, where most of the stars are.  By far the largest threat are the occasional radio galaxy episodes that can fill these galaxies with high levels of CRs.  These galaxies may alternate between epochs of moderate or high traversability (at least in the denser inner regions) and non-traversability.

\emph{Red nuggets} -- the very compact high mass fast rotators commonly found in the early Universe \citep{Trujillo06-RedNugget}, although a small number of examples are known at $z \sim 0$ \citep{Trujillo14}.  These are very compact high mass ETGs.  The stars are packed densely and have a high velocity dispersion, greatly favoring traversability.  X-ray studies of one nearby red nugget revealed a large halo of hot gas, which becomes quite dense ($\sim 0.3\ \cm^{-3}$) in the inner kiloparsec, where most of the stars reside \citep{Werner18}.  Thus, gas could impede traversability somewhat in these galaxies, possibly offset by the higher density of destinations and relative lack of dust and CRs.  The hot gas may fuel radio galaxy episodes.    

\emph{Globular clusters} -- although not galaxies, the conditions are different enough from their host galaxy to warrant special mention.  Globular clusters can be extremely dense, so interstellar journeys are very short \citep{DiStefano16}.  Globular clusters do have an ISM, low density ($\sim 0.1\ \cm^{-3}$ in 47 Tucanae) and in most cases ionized, even though it is not clear why there is not much more \citep{Freire01,vanLoon06}.  Small amounts of dust probably accompanies the gas. Some shine in gamma-rays with the glow of their pulsars, which might release high energy CR electrons and positrons into the ISM, but otherwise there should be few CRs \citep{Abdo10-Globulars}.  If they are near a SFG like the Milky Way, they could be exposed to its CRs as they escape the galaxy, but very little is known about CRs far in the halo (compare the very different models of \citealt{Breitschwerdt91} and \citealt{Feldmann13}).  Globular clusters are very likely to be super-traversable. 

\emph{Galaxy clusters} -- the largest dark matter halos, of which the member galaxies are analogous to satellites.  Galaxy clusters have their own diffuse population of stars, the intracluster light (ICL), which is also an outer halo of their central bright galaxy \citep{Kluge20}.  These move with tremendous velocity dispersions (approaching $1,000\ \kms$).  The stellar density is very low, however, with stars separated by $\sim 100\ \pc$.  In addition, galaxy clusters have an intracluster medium (ICM), an all-pervading low density $10^8\ \Kelv$ plasma.  Its density peaks towards the cluster center \citep{Cavaliere13}.  The ICM does sometimes contain CRs, visible as radio ``halos'' or ``relics'' \citep{vanWeeren19}, but the CR energy density is constrained to be less than about one-tenth the Milky Way level (based on a CR/thermal pressure ratio $\la 1\%$ in \citealt{Arlen12,Ackermann14-Clusters}). \editOne{ICM magnetic fields of a few microgauss are present, rising to tens of microgauss in the hearts of certain cooling core clusters \citep{vanWeeren19}; these are turbulent fields, but the fluctuation scale is so large that they should be easy to account for on interstellar journeys.} The traversability per unit length is favorable compared to the Milky Way, but the distance between intracluster stars increases exposure to the ICM and makes it much more likely that internal hazards disrupt a journey.

\subsection{Summary}
Overall, and perhaps not too surprisingly, quiescent galaxies appear to have the best prospects for interstellar travel, significantly better than the Milky Way.  Globular clusters and compact ellipticals likely are super-traversable: their ISM is minuscule and the density of stars is extremely high.  Larger compact ellipticals like M32 have the added advantage of relatively high stellar velocity dispersions, mixing stellar populations thoroughly and bringing a stream of new potential destinations to each sun.  But even the majority of ETGs are plausibly super-traversable most of the time, with relatively dilute ISMs and dense stellar populations near their cores. The centers of the largest dwarf spheroidals may also be super-traversable, but smaller ones could be sub-traversable because of the low $\rhostar$. SFGs are expected to be relatively poor in comparison, although dwarf irregulars with sufficiently high $\rhostar$ are somewhat favorable compared to the Milky Way.  Starbursts and the SFGs of the distant past -- including the ancient Milky Way -- were subtraversable from an ISM standpoint, filled with hazards to interstellar travel. The dense stellar populations of nuclear starbursts, like those in the cores of other galaxies, may aid traversability by shortening interstellar trip. 

\section{Traversability and SETI strategy}

Traversability has implications for SETI, as it potentially regulates the development of the most technologically powerful societies detectable from the greatest distances.  If self-propagating interstellar settlement possible, a single society originating on a single planet has the potential to spread across an entire galaxy, developing into \emph{billions} within a hundred million years.  Moreover, these societies can resettle worlds where ETI society goes extinct, as long as the lifespan is not too short \citep{CarrollNellenback19,Dosovic19}.  From our perspective, the entire collection of societies would then have an unlimited lifespan.

Two points follow if this is true.  First, in a collection of millions of societies with diverse values and motivations, there is a greater chance that \emph{someone} at any given time will be interested in maintaining a beacon detectable from intergalactic distances.  Even if these beacons are maintained for only a short period (say, a few years) in any given system, another society could step up with their own, leading to a steady state population of beacons.  In contrast, if the Milky Way is untraversable, then there may only be a few thousand ETIs present at any given time, most of whom may not be interested in operating a long-term expensive beacon \citep[c.f.,][]{Sagan73,Bates78}.

Second, interstellar propagation might nullify all but the strictest habitability concerns.  It simply takes one ETI to arise and successfully spread across the galaxy to seed millions more.  So even though low metallicity negatively affects habitability in dwarf spherodials, and there conceivably are other issues with ETGs, just a few habitable planets may suffice if technological intelligence are ubiquitous.  In fact, this may not matter too much, as compact ellipticals and larger ETGs galaxies are both proposed to be favorable for habitability \citep{Dayal15,Stojkovic19}.  It might also be the case that powerful ETIs that can manage intergalactic journeys might deliberately migrate to uninhabited but traversable galaxies like dwarf spheroidals, perhaps specifically for the isolation. The cores of galaxies are plausibly also favored for SETI despite the high rate of potentially dangerous events, because their high stellar densities enhance traversability \citep[c.f.,][]{Gajjar21}.

Traversability adds another factor to the discussion of the Fermi Paradox.  One proposed partial solution is that the Milky Way was less habitable billions of years ago.  \citet{Cirkovic08} argues that this could have lead to a ``phase transition'' between an epoch when sterilization events inhibit the evolution of ETIs and the recent era when ETIs are finally appearing and potentially spreading.  The Milky Way was also less traversable in the distant past, with a star-formation rate of about five to ten times its present rate \citep{Snaith15}, suggesting that even if ETIs evolved billions of years ago, the stronger radiation could have confined them to their homeworlds.  However, this effect should not be overstated. Even leaving aside the possibility of shielding, the star-formation rate of the Milky Way has not been a steady decline, but fluctuates significantly over cosmic time, with an intense episode about 10 Gyr ago \citep{Gallart19}, perhaps even ceasing for about a billion years at one point about 8 Gyr ago \citep{Snaith15,Haywood16}.  Thus, the Milky Way could have been more traversable than at present during low SFR episodes billions of years ago.

Large slow rotator galaxies could have an interesting analog of the sterilization events invoked in galactic habitability discussion.  They may have galaxy-wide \emph{lockdown} events during radio galaxy episodes.  At worst, these lockdowns could result in the extinction of all the societies as they are unable to resettle worlds that have fallen.  Even if the ETIs on each world are effectively immortal, they may have to alter their behavior, relying more on remote communication through radio or optical broadcasts instead of direct visitation.

\editOne{The galactic environment could shape which technologies are used for interstellar travel, favoring those with fewer traversability hazards.  This would suggest a kind of negative technological feedback that compensates for traversability differences.  Yet it also suggests that the details of the ``propagation'' of ETIs through space could vary from galaxy to galaxy, with implications for models of this diffusion \citep[c.f.,][]{Jones76,CarrollNellenback19}.  Traversability also plays a more conventional role for SETI even if settlement over interstellar distances is impractical.  There is a long-standing debate about whether ETIs would contact their neighbors by electromagnetic broadcasts or by instead sending physical artifacts or probes \citep[e.g.,][]{Bracewell60,Rose04}.  Traversable galaxies could inspire their ETIs to develop probe technology instead of electromagnetic broadcasts without increasing the number of ETIs if crewed interstellar flight is impractical, depriving us of technosignatures. However, ETIs who are deliberately trying to signal their extragalactic peers presumably would employ contact methods that are readily detectable at great distance.}

Finally, traversability can introduce percolation effects in the propagation of ETIs, particularly in SFGs.  The turbulent, multiphase ISM will make travel easier for some worlds and along some routes than others.  As with the older culturally-driven percolation hypothesis, though, mixing within galaxies will open up new paths over time, preventing permanent containment of propagation.

\section{Summary and Outlook}

Galactic traversability describes how difficult interstellar travel is in a galaxy.  Traversability is a potential limiting factor in the development of galaxy-wide societies that would stand out from afar.  In this paper, I motivated both the stellar population and the ISM of a galaxy as contributing or hampering starfaring.  After reviewing our current knowledge about various types of galaxies, I concluded that quiescent galaxies are on the whole likely to be more traversable than the Milky Way, although they are not entirely free of hazards.  Compact ellipticals like M32 are judged to have the highest traversability, being dense, high dispersion stellar systems with very little ISM. While there is some overlap between habitable galaxies and traversable galaxies, particularly for medium/large quiescent galaxies and possibly compact ellipticals, some potentially traversable galaxies like larger dwarf spheroidals have poor habitability.  As long as a single starfaring ETI evolves in (or reaches) these low-habitability galaxies, these galaxies may become populated anyway and could be good SETI targets.

Most SETI programs have focused on the Milky Way, but targeted extragalactic campaigns have covered the SMC (\citealt{Shostak96}), M31, and M33 \citep{Gray17}.  None of these galaxies are obviously super-traversable, but they all have some favorable aspects: the SMC has less dust, and all three have lower CR backgrounds \citep{Abdo10-SMC,Ackermann17}.  The most comprehensive targeted campaign of other galaxies is the Breakthrough Listen nearby galaxies survey.  It specifically targets several morphological classes of galaxies, including twenty dwarf spheroidals and forty-three large early-type galaxies (many of the latter in the Virgo Cluster, and thus likely to be gas poor), thus providing a relatively large sample of super-traversable galaxies \citep{Isaacson17}.  Several more are in the Breakthrough Listen Exotica Catalog, along with some globular clusters and galaxy clusters \citep{Lacki20}.  M32, the only proper compact elliptical in the Local Group, partly overlaps the \citet{Gray17} survey of M31 and will be covered by the Breakthrough Listen survey of M31 \citep{Li20}.

Traversability can also be a consideration when searching for Kardashev Type III societies, as it may be a limiting factor in their development.  Some searches for these ETIs have focused on star-forming spiral galaxies \citep[e.g.,][]{Zackrisson15,Garrett15}, but quiescent galaxies may have better prospects.  The Fundamental Plane of ETGs can aid the search for cloaked stellar populations \citep{Annis99}.  Waste heat surveys should have detected large cloaked quiescent galaxies \citep{Griffith15}, but limits on cloaked dwarf galaxies and globular clusters are currently weak \citep[e.g.,][]{Lacki16-K3}; deeper surveys of galaxy clusters could be useful.

Future avenues for theoretical research include study of percolation effects due to ISM conditions, and the effects of radio galaxy lockdowns in large slow rotator galaxies.  Of course, the greatest theoretical uncertainty is whether traversability applies at all, or if the concerns are overcome through adaptation.  For example, large ``worldships'' with extensive shielding may be immune to dust and CRs.  Some of the ``hazards'' could even be turned into advantages -- dense gas could be harvested for propellant, and ionized gas can be used to brake with a magnetic sail.  

\editOne{Although most of an interstellar craft's time is spent far from stars, some technologies are subject to stellar-level traversability effects.  The different luminosities of stellar objects affects the performance of solar sail type craft \citep{Lingam20}. \citet{Heller17} explored how stars of different luminosity and mass could be used for braking or maneuvering.  Different types of stars have different stellar wind parameters, varying the performance of magnetic and electric sails.  Interplanetary traversability no doubt is also affected by the architecture of planetary systems, with shorter travel times in compact systems commonly found around red dwarfs (\citealt{Swift13}; see also similar considerations about panspermia in \citealt{Lingam17-TRAPPIST}).}

If the presence of an ISM is an important factor for traversability, the forecast for starfarers is bright.  As gas and star formation continues to decline, the ISM of galaxies should become more traversable.  This is an erratic process, with bursts and respites in star formation over billions of years.  Over tens of billions of years, the Local Group should coalesce into a single elliptical galaxy.  Depending on whether the Milky Way's descendant retains enough of a hot atmosphere to fuel radio galaxy episodes, interstellar space could be cleared of hazards. Of course, if ETIs favor using RAIR propulsion, traversability decreases with time; either a new propulsion method would be needed in this distant epoch or interstellar travel will cease.  The Galaxy's $\vstar$ should increase through secular processes and then the merger with M31, although $\rhostar$ could decrease to roughly compensate it in terms of traversability.  Ironically, even if interstellar travel becomes easier, intergalactic travel is becoming harder due to the accelerating cosmic expansion \citep{Heyl05}.  Trillions of years from now, the descendants of galaxies in today's galaxy clusters may still be more traversable than the fossil Local Group, simply because there will be more places to go in their cosmic horizon.

\section*{Conflict of Interest}
The author reports no conflicts of interest.

\ack[Acknowledgements]{I thank the Breakthrough Listen program for their support. Funding for \emph{Breakthrough Listen} research is sponsored by the Breakthrough Prize Foundation.\footnote{{\url{https://breakthroughprize.org/}}}

In addition, I acknowledge the use of NASA's Astrophysics Data System and arXiv for this research. \editOne{I thank the referees for reading and commenting on the manuscript.}}

\end{document}